\documentclass[showpacs,pre]{revtex4}
\usepackage{graphicx}
\begin{document}
\title{Instability of synchronized motion in nonlocally coupled neural oscillators}
\author{Hidetsugu Sakaguchi}
\affiliation{Department of Applied Science for Electronics and Materials,\\
Interdisciplinary Graduate School of Engineering Sciences,\\
Kyushu University, Kasuga, Fukuoka 816-8580, Japan}
\begin{abstract}
 We study nonlocally coupled Hodgkin-Huxley equations with excitatory and inhibitory  synaptic coupling. We investigate the linear stability of the synchronized solution, and find numerically various nonuniform oscillatory states such as chimera states, wavy states, clustering states, and spatiotemporal chaos as a result of the instability.
\end{abstract}
\maketitle
\section{Introduction}
Synchronization of neural oscillators is considered to play an important role in various functions such as visual information processing, sleeping, and memory in the brain \cite{rf:1,rf:2,rf:3}.  The synchronization of neural oscillators has been theoretically and numerically studied by many authors \cite{rf:4,rf:5,rf:6,rf:7,rf:8}.  The conditions of the synchronization in the neural oscillators were intensively studied, but it is not well known what happens after the synchronized motion becomes unstable.   Synchronization is not always desirable; several neurological diseases such as Parkinson's disease and epilepsy are caused by synchronized firing of neural oscillators. The control and prediction of  synchronized motion are related to therapies for the diseases.  
A therapy for Parkinson's disease is external electric stimulation at high frequencies, called deep brain stimulation \cite{rf:9}. Time series analyses of electroencephalograms were studied to predict an epilepsy seizure beforehand \cite{rf:10}.  Some drugs for neurological diseases are interpreted to modify synaptic currents, change the interaction among neural oscillators, and make anomalously synchronized states change into asynchronous states.
 
  The stability of the synchronized solution depends on the details of the model equations of neural oscillators. However, there is some general tendency that excitatory coupling induces  synchrony and inhibitory coupling leads to  antisynchrony (in which the phase difference between two oscillators is $\pi$), if the response time of synaptic couplings is rather small.   However, if the response time of synaptic couplings is slower,  the situation is reversed. The synchronous state is stable for inhibitory coupling, and the synchronous state becomes unstable for excitatory coupling.   The response time or the delay time is an important parameter for the stability of the synchronized motion.    
The stability of the synchronous state has been studied mainly in coupled two-neuron systems or in uniformly or randomly coupled systems of many neurons.  On the other hand, Kuramoto and collaborators found various complicated solutions in nonlocally coupled complex Ginzburg-Landau equations and nonlocally coupled phase oscillators \cite{rf:11,rf:12}. We studied nonlocally coupled noisy integrate-and-fire models using the Fokker-Planck equation and found traveling wave solutions \cite{rf:13}. In this paper, we study the instability of the synchronized solution in nonlocally coupled Hodgkin-Huxley equations by changing the response time of the synaptic coupling. 
\section{Nonlocally coupled Hodgkin-Huxley equation} 
The Hodgkin-Huxley equation is a fundamental model equation for the firing of neurons. The equation was originally proposed for the giant axon in a squid 
\cite{rf:14}. Generalized Hodgkin-Huxley type equations are used for various neurons. These are coupled differential equations for the membrane potential $V$ and the auxiliary variables $m,n,h$, which are related to the conductance of ion currents through the Na$^+$ and K$^+$ channels and a leak current. The Hodgkin-Huxley equation is written as \cite{rf:14} 
\begin{eqnarray}
C_m\frac{dV}{dt}&=&g_{{\rm Na}}m^3h(E_{{\rm Na}}-V)+g_{\rm K}n^4(E_{\rm K}-V)+g_L(E_{\rm L}-V)+I_0+I_s,\nonumber\\
\frac{dm}{dt}&=&\alpha_m-(\alpha_m+\beta_m)m,\nonumber\\
\frac{dh}{dt}&=&\alpha_h-(\alpha_h+\beta_h)h,\nonumber\\
\frac{dn}{dt}&=&\alpha_n-(\alpha_n+\beta_n)n,\nonumber\\
\alpha_m&=&\frac{0.1(V+40)}{1-\exp(-(V+40)/10)},\;\;\beta_m=4\exp(-(V+65)/18),\nonumber\\
\alpha_h&=&0.07\exp(-(V+65)/20),\;\;\beta_h=\frac{1}{1+\exp(-(V+35)/10)},\nonumber\\
\alpha_n&=&\frac{0.01(V+55)}{1-\exp(-(V+55)/10)},\;\;\beta_n=0.125\exp(-(V+65)/80),
\end{eqnarray}
where $C_m=1,\,g_{{\rm Na}}=120,\,g_{{\rm K}}=36,\,g_{{\rm L}}=0.3,\,E_{{\rm Na}}=50, \,E_{{\rm K}}=-77$ and $E_{{\rm L}}=-54.4$, $I_0$ denotes the external current stimulus, and $I_s$ is the synaptic current. The unit of electric potential is the millivolt and the unit of time is the millisecond. We propose a time evolution equation for the synaptic current $I_s$ in a one-dimensional system as 
\begin{equation}
\tau\frac{dI_s}{dt}=\int_{-\infty}^{\infty}g(x-x^{\prime})F(V(x^{\prime}))dx^{\prime}-I_s,
\end{equation}
where $g(x)$ represents a coupling function of the nonlocal synaptic interaction, $\tau$ is the response time, and $F(V)$ is the response function of the synaptic current to the membrane potential. In this paper, we assume a function 
\begin{eqnarray}
F(V)&=&\gamma(V-V_0) \;\;\;\;{\rm for}\;\;V>V_0,\nonumber\\
&=&0 \;\;\;\;\;\;\;\;\;\;\;\;\;\;\;\;\;\;{\rm for}\;\;V<V_0,
\end{eqnarray}
where the parameters $V_0=-50$ and $\gamma=0.01$ are used in this paper. 
As a coupling function $g(x)$, we use an exponential function $g(x)=g_1\alpha/2\exp(-\alpha|x|)$ or a sum of two exponential functions $g(x)=g_1\alpha_1/2\exp(-\alpha_1|x|)+g_2\alpha_2/2\exp(-\alpha_2|x|)$ for the sake of simplicity.
 For synchronized solutions, $V(x,t),m(x,t),h(x,t)$ and $n(x,t)$ do not depend on $x$;  
then the uniform solutions $V_0(t),m_0(t),h_0(t)$ and $n_0(t)$ obey
the same equation (1), but the synaptic current $I_{s0}$ obeys 
\begin{equation}
\tau\frac{dI_{s0}}{dt}=g_0F(V_0)-I_s,
\end{equation}
where $g_0=\int_{-\infty}^{\infty}g(x)dx$. 

\section{Instability of synchronized motion in excitatory and inhibitory systems} 
First, we investigate the linear stability of the uniform solution for the nonlocally coupled Hodgkin-Huxley equation. Small perturbations with wave number $k$ around the uniform solution: $\delta V=V_k\exp(ikx), \delta m=m_k\exp(ikx),\delta h=h_k\exp(ikx),\delta n=n_k\exp(ikx)$ and $\delta I_s=I_{sk}\exp(ikx)$ obey
\begin{eqnarray}
C_m\frac{dV_k}{dt}&=&g_{{\rm Na}}\{3m_0^2m_kh_0(E_{{\rm Na}}-V_0)+m_0^3h_k(E_{{\rm Na}}-V_0)-m_0^3h_0V_k\}\nonumber\\
&+&g_{{\rm K}}4n_0^3n_k(E_{\rm K}-V)+g_{{\rm K}}n_0^4(-V_k)+g_{{\rm L}}(-V_k)+I_{sk},\nonumber\\
\frac{dm_k}{dt}&=&\frac{\partial \alpha_m}{\partial V}V_k-\left(\frac{\partial \alpha_m}{\partial V}+\frac{\partial \beta_m}{\partial V}\right )m_0V_k-(\alpha_m+\beta_m)m_k,\nonumber\\
\frac{dh_k}{dt}&=&\frac{\partial \alpha_h}{\partial V}V_k-\left(\frac{\partial \alpha_h}{\partial V}+\frac{\partial \beta_h}{\partial V}\right )h_0V_k-(\alpha_h+\beta_h)h_k,\nonumber\\
\frac{dn_k}{dt}&=&\frac{\partial \alpha_n}{\partial V}V_k-\left(\frac{\partial \alpha_n}{\partial V}+\frac{\partial \beta_n}{\partial V}\right )n_0V_k-(\alpha_n+\beta_n)n_k,\nonumber\\
\tau\frac{dI_{sk}}{dt}&=&g_k\frac{\partial F(V)}{\partial V}V_k-I_{sk},
\end{eqnarray}
where $g_k$ is the Fourier transform of $g(x)$. If the uniform solution $V_0,m_0,h_0,n_0$, and $I_{s0}$ is time periodic, the small perturbation increases or decays periodically. We can calculate numerically the linear growth rate of the norm $N=\sqrt{V_k^2+m_k^2+h_k^2+n_k^2+I_{sk}^2}$. 
\begin{figure}[htb]
\begin{center}
\includegraphics[height=4.5cm]{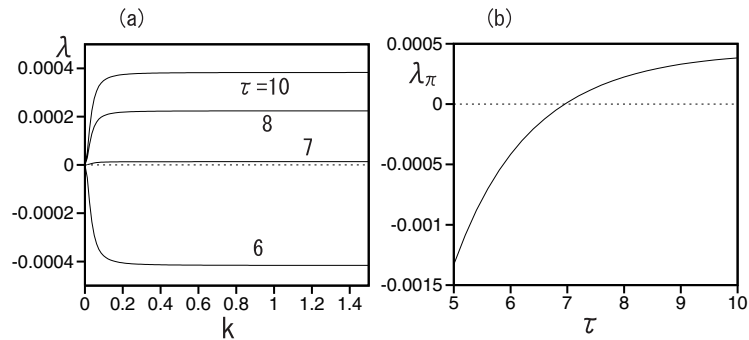}
\end{center}
\caption{(a) Linear growth exponent $\lambda$ for the perturbation with wavenumber $k$ for the excitatory coupling $g(x)=1.8\exp(-0.03|x|)$.
(b) $\lambda$ for $k=\pi$ as a function of the response time $\tau$.}
\label{f1}
\end{figure}
\begin{figure}[htb]
\begin{center}
\includegraphics[height=4cm]{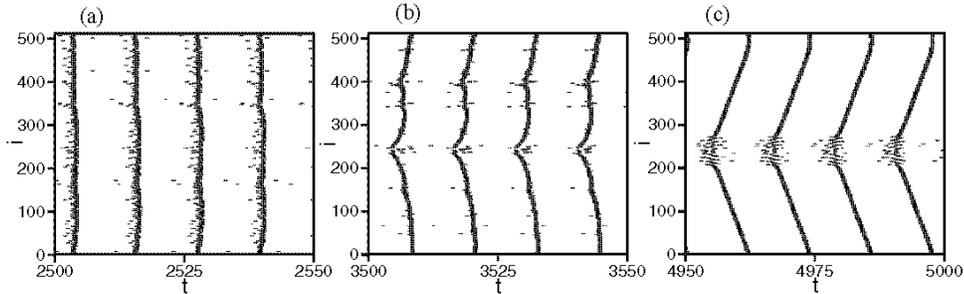}
\end{center}
\caption{Raster plots of firing neurons at $\tau=8.5$ for (a) $2500<t<2550$, (b) $3500<t<3550$ and (c) $4950<t<5000$. }
\label{f2}
\end{figure}
\begin{figure}[htb]
\begin{center}
\includegraphics[height=4cm]{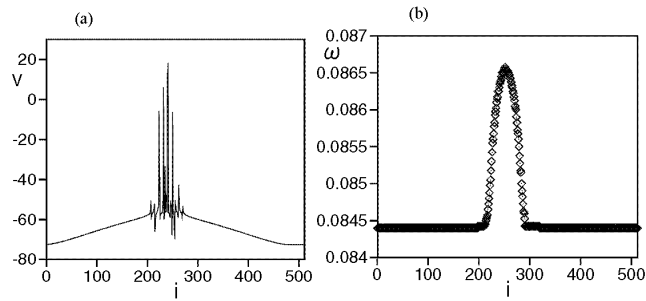}
\end{center}
\caption{(a) Snapshot of the membrane potential $V(i)$ at $\tau=8.5$. (b) Frequency profile $\omega(i)$. }
\label{f3}
\end{figure}
Numerical simulations are performed using the Runge-Kutta method. The system system size is $L=512$ and the grid sizes $\Delta t=0.005$ and $\Delta x=1$, and periodic boundary conditions are imposed. That is, 512 oscillators are set on a one-dimensional circle of length 512.  Figure 1(a) displays the average linear growth exponent  $\lambda=\lim (1/T){\rm ln} [N(T)/N(0)]$ of the norm for large $T$ as a function of $k$ for an excitatory coupling $g(x)=1.8\exp(-0.03|x|)$ and the uniform input $I_0=15$. The four curves correspond to four response times $\tau=6,7,8$ and 10. The synchronized state is stable at $\tau=6$ and unstable at $\tau=7$ for all wave numbers.  Figure 1(b) displays the linear growth rate $\lambda$ for $k=\pi$. The instability occurs at $\tau=6.7$. 
Figures 2(a),(b), and (c) display raster plots of firing neurons for some time intervals (a) $2500<t<2550$, (b) $3500<t<t<3550$, and (c) $4950<t<5000$ at $\tau=8.5$. In the raster plots, positions of neural oscillators satisfying the firing condition $V(i,t)>0$  are plotted with dots at every time interval 0.25 (ms). 
The initial condition  is a uniform state with small random perturbation. 
The uniform state is unstable and the random perturbation enlarges in time as shown in Fig.~2(a). 
Discontinuities appear in the profile of membrane potential $V(i)$, because the nonlocal coupling does not prohibit the discontinuity in contrast to the diffusion couplings. There appear mainly two discontinuous regions at $i\sim 230$ and $i\sim 400$ in Fig.~2(b). One discontinuous region survives at $t=5000$ as seen in Fig.~2(c). The discontinuous region plays a role of a pacemaker and excitation pulses  propagate toward both sides. Figure 3(a) displays a profile of the membrane potential $V$ at $\tau=8.5$. The membrane potential $V$ is discontinuously distributed in $200<i<280$.  The time evolution of $V$ is almost periodic, and the synaptic current $I_s$ changes in time almost periodically  between 6 and 13.  
Figure 3(b) displays the frequency profile $\omega(i)$ for the neural oscillators. The frequency of the oscillation in the discontinuous region is slightly faster than that in the continuous region, and it changes smoothly in space. This  is similar to chimera states studied in Ref.[12,15].  Although they found the chimera states in  conditions where the uniform state is stable, our chimera state has  spontaneously appeared as a result of the instability of the uniform state.       

\begin{figure}[htb]
\begin{center}
\includegraphics[height=4.5cm]{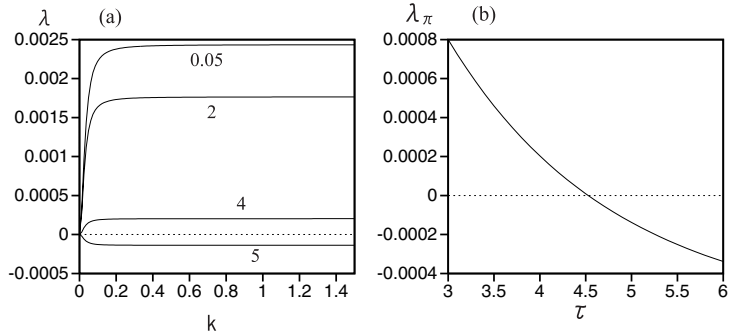}
\end{center}
\caption{(a) Linear growth exponent $\lambda$ as a function of wave number $k$ for the inhibitory coupling $g(x)=-1.8\exp(-0.03|x|)$.
(b) Linear growth exponent $\lambda$ at $k=\pi$ as a function of the response time $\tau$.}
\label{f4}
\end{figure}
\begin{figure}[htb]
\begin{center}
\includegraphics[height=6cm]{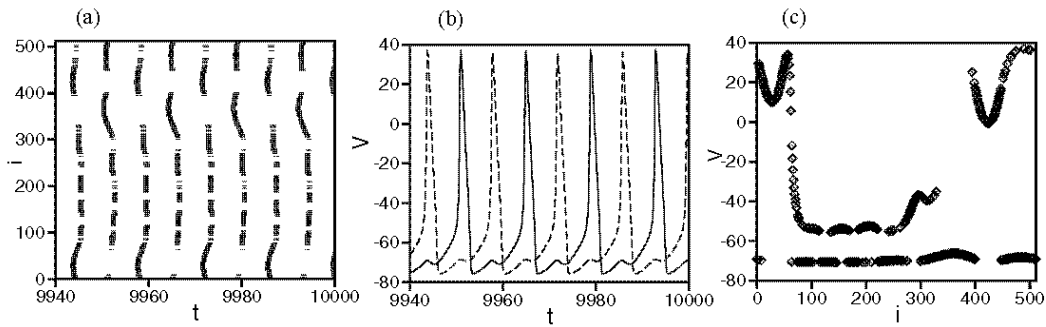}
\end{center}
\caption{(a) Raster plot for $\tau=4$. (b) Time evolution of $V(i)$ for $i=1$ (solid curve) and $i=5$ (dashed curve. (c) Snapshot of the membrane potential $V(i)$ at $t=9985$.}
\label{f5}
\end{figure}
Next, we consider an inhibitory system. The coupling function $g(x)$ is assumed as $g(x)=-1.8\exp(-0.03|x|)$, and the uniform input is $I_0=15$. Figure 4(a) displays the average linear growth rate $\lambda$ as a function of wavenumber $k$ for $\tau=0.05,2,4$, and 5. Figure 4(b) displays the average linear growth rate $\lambda$ for $k=\pi$ as a function of $\tau$. The uniformly synchronized state is unstable for $\tau<4.4$.
For the inhibitory coupling, the uniform state changes into a spatially modulated clustering state. Figure 5(a) displays a raster plot of firing neurons at $\tau=4$. 
Figure 5(b) displays time evolutions of $V(i)$ for $i=1$ and 5. 
The time evolution is almost periodic, and it is clearly seen that the phase difference between the two oscillators $i=1$ and 5 is about $\pi$.
 Figure 5(c) displays  a profile of the membrane potential $V$ at $t=9985$. 
The phase of each oscillator in each cluster is not uniform, but changes slowly in space. However, this phase profile is periodically recovered as seen in the raster plots in Fig.~5(a).  The difference of the membrane potential $V(i)$ between two oscillators inside the same cluster changes periodically but recovers the same value after one period of the oscillation. That is, the phase profile is random but it is frozen like a glassy state. 

\begin{figure}[htb]
\begin{center}
\includegraphics[height=4.cm]{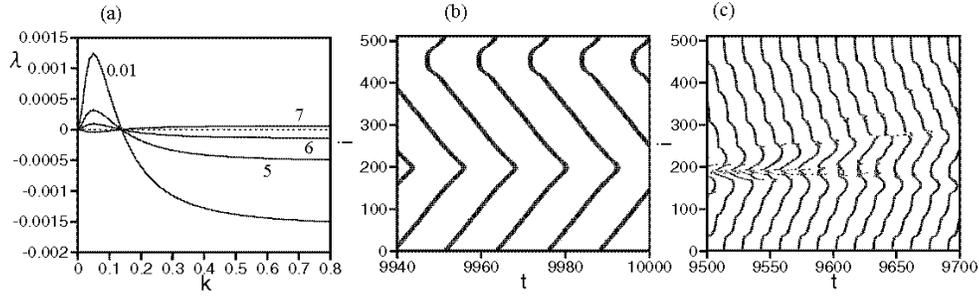}
\end{center}
\caption{(a) Linear growth exponent $\lambda$ for the perturbation with wave number $k$ for the coupling $g(x)=-1.8\exp(-0.03|x|)+12\exp(-0.12|x|)$.
(b) Raster plot to display firing neurons at $\tau=0.01$. (c) Raster plot at $\tau=5$. }
\label{f6}
\end{figure}
\begin{figure}[htb]
\begin{center}
\includegraphics[height=9cm]{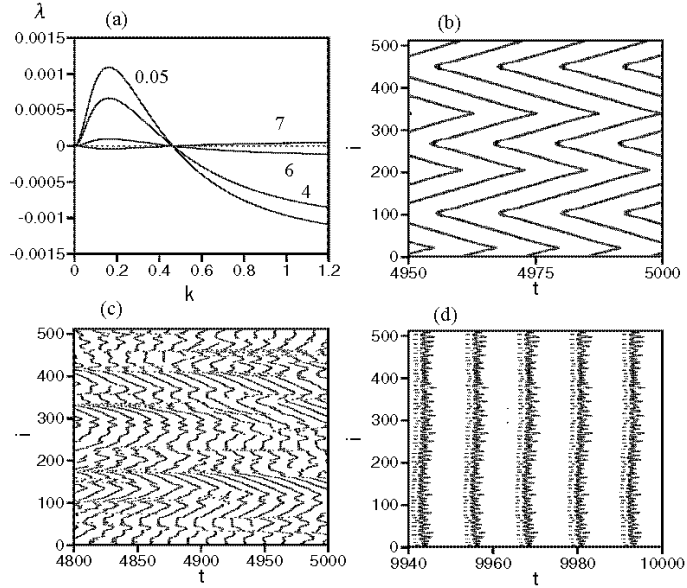}
\end{center}
\caption{(a) Linear growth exponent $\lambda$ for the perturbation with wave number $k$ for the inhibitory coupling $g(x)=-6\exp(-0.03|x|)+40\exp(-0.12|x|)$.
(b) Raster plot to display firing neurons at $\tau=0.05$. (c) Raster plot at $\tau=5$. (d) Raster plot at $\tau=8$.}
\label{f7}
\end{figure}

\section{Competition of excitatory and inhibitory interactions}
We can consider a more complicated system with excitatory and inhibitory interactions.  We study the case of $g(x)=-1.8\exp(-0.03|x|)+12\exp(-0.12|x|)$. The uniform input $I_0=15$. The interaction is excitatory in the short range and inhibitory in the long range. This type of interaction is often used in artificial neural network models. The uniform input is assumed to be $I_0=15$. 
Figure 6(a) displays the average linear growth rate $\lambda$ as a function of wave number $k$ for $\tau=0.01,5,6,$ and 7.  The growth rate $\lambda$ becomes zero at $k=k_c=0.139$ for every $\tau$, where $k_c$ is a solution of $g_k=-1.8\cdot 0.03/(k^2+0.03^2)+12\cdot 0.12/(k^2+0.12^2)=40=g_0$, where $g_k$ is the Fourier transform of $g(x)$. For $\tau<6.4$, the growth rate $\lambda>0$ for smaller $k$ with $k<k_c$, and $\lambda<0$ for larger $k$ with $k>k_c$. On the other hand, in the case of $\tau>6.4$, $\lambda<0$ for $k<k_c$, and $\lambda>0$ for $k>k_c$.
There is a peak in the curve of $\lambda(k)$ at a finite $k$, which corresponds to the most unstable mode.
The existence of the characteristic wavelength is a result of the competition of the excitatory and inhibitory interactions.  
Figure 6(b) displays a raster plot of firing neurons at $\tau=0.01$.  
A pacemaker region appears near $i=450$.  
Traveling pulses are sent out regularly toward both sides from the pacemaker region.  The spatial continuity is maintained for this coupling and therefore the frequency distribution is uniform in contrast to the chimera state.
Figure 6(c) displays a raster plot of firing neurons at $\tau=5$. The traveling wave state becomes unstable, spatial modulations grow up  and a spatiotemporal chaos appears. Discontinuities sometimes appear in the raster plot as a result of the spatiotemporal chaos. 

Figure 7 shows a numerical result for a system with another coupling function $g(x)=-6\exp(-0.1|x|)+40\exp(-0.4|x|)$.  In this coupling function, the critical wave number is $k_c=0.464$.  The critical value of $\tau$ for the instability is almost the same $\tau_c=6.4$ as the case shown in Fig.~7.  The growth rate takes a maximum at a finite parameter $k\sim 0.16$ for $\tau\sim 6.4$. Figure 7(b) displays a raster plot of firing neurons at $\tau=0.05$. Regular traveling waves state appear. There are three pacemakers. Because the ratio of the typical wave number for the case of Figs.~6 and 7 is $0.139/0.464=0.3$, it is natural that the number of pacemakers is three times the case of Fig.~6.  Figure 7(c) displays a raster plot of firing neurons at $\tau=5$. Spatial modulations grow up and a spatiotemporal chaos appears. Discontinuities sometimes appear, and they induce phase slips. As a result, the frequency profile $\omega(x)$ is almost uniform, but randomly distributed around the uniform value. That is, the complete entrainment is broken by the phase slips. Figure 7(d) displays a raster plot of firing neurons at $\tau=8$. Perturbations with small wavelength increase as seen in Fig.~7(d), because the linear instability occurs only for large wave number $k>k_c$. 
  
\section{Instability of synchronized motion in two-layer models}
In neural systems, the characteristic of the interaction (that is, excitatory coupling or inhibitory coupling) is usually determined by the property of the presynaptic neuron. This is called Dale's principle. 
The interaction considered in the previous section is not so suitable in realistic neural systems.
\begin{figure}[htb]
\begin{center}
\includegraphics[height=8cm]{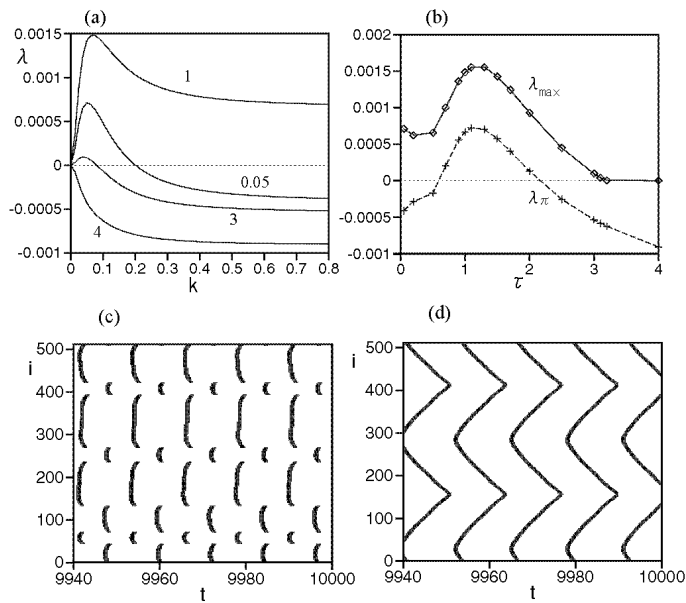}
\end{center}
\caption{(a) Linear growth exponent $\lambda$ for the perturbation with wavenumber $k$ for the two-layer model with coupling functions $g_1(x)=5.4\exp(-0.12|x|), g_2(x)=-0.9\exp(-0.03|x|),;g_3(x)=3.6\exp(-0.12|x|)$ and $g_4(x)=0$.
(b) Maximum value of $\lambda$ (solid curve) and $\lambda_{\pi}$ (dashed curve) for the wavenumber $k=\pi$ as a function of $\tau$. (c) Raster plot to display firing neurons in the first layer at $\tau=0.05$. (d) Raster plot to display firing neurons in the first layer at $\tau=2$.}
\label{f8}
\end{figure}
\begin{figure}[htb]
\begin{center}
\includegraphics[height=4.5cm]{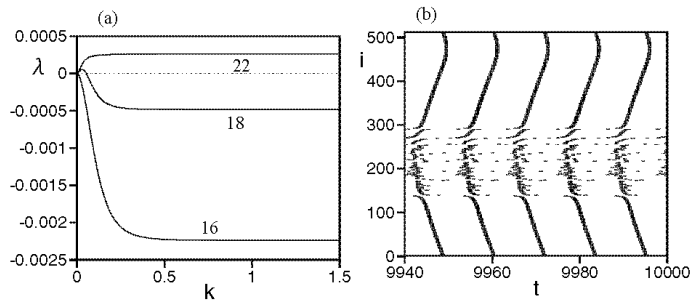}
\end{center}
\caption{a) Linear growth exponent at $\tau=22,18$ and 16 for the coupling functions $g_1(x)=1.35\exp(-0.03|x|), g_2(x)=-3.6\exp(-0.12|x|),g_3(x)=3.6\exp(-0.12|x|)$ and $g_4(x)=0$.(b) Raster plot of the neurons in the first layer at $\tau=22$
}
\label{f9}
\end{figure}
We can consider a two-layer model, where excitatory neurons in the first layer  interact with inhibitory neurons in the second layer.  
The Hogdkin-Huxley equations for the two-layer model are written as
\begin{eqnarray}
C_m\frac{dV_1}{dt}&=&g_{{\rm Na}}m_1^3h_1(E_{{\rm Na}}-V_1)+g_{\rm K}n_1^4(E_{\rm K}-V_1)+g_L(E_{\rm L}-V_1)+I_{01}+I_{s1}+I_{s2},\nonumber\\
C_m\frac{dV_2}{dt}&=&g_{{\rm Na}}m_2^3h_2(E_{{\rm Na}}-V_2)+g_{\rm K}n_2^4(E_{\rm K}-V_2)+g_L(E_{\rm L}-V_2)+I_{02}+I_{s3}+I_{s4},\nonumber\\
\tau_1\frac{dI_{s1}}{dt}&=&\int_{-\infty}^{\infty}g_1(x-x^{\prime})F(V_1(x^{\prime}))dx^{\prime}-I_{s1},\nonumber\\
\tau_2\frac{dI_{s2}}{dt}&=&\int_{-\infty}^{\infty}g_2(x-x^{\prime})F(V_2(x^{\prime}))dx^{\prime}-I_{s2},\nonumber\\
\tau_3\frac{dI_{s3}}{dt}&=&\int_{-\infty}^{\infty}g_3(x-x^{\prime})F(V_1(x^{\prime}))dx^{\prime}-I_{s1},\nonumber\\
\tau_4\frac{dI_{s4}}{dt}&=&\int_{-\infty}^{\infty}g_4(x-x^{\prime})F(V_2(x^{\prime}))dx^{\prime}-I_{s4},
\end{eqnarray}
where $V_1$ and $V_2$ are membrane potentials of neurons in the first and second layers, and $I_{s1},I_{s2},I_{s3}$, and $I_{s4}$ are, respectively, synaptic currents by the interaction inside the first layer, from the second layer to the first layer, from the first layer to the second layer, and inside the second layer.
By Dale's principle, we assume that $g_1>0,\;g_2<0,\;g_3>0$ and $g_4<0$. 
 The interaction functions are assumed to be $g_1(x)=5.4\exp(-0.12|x|), g_2(x)=-0.9\exp(-0.03|x|), g_3(x)=3.6\exp(-0.12|x|)$ and $g_4(x)=0$. The response times $\tau_1,\tau_2,\tau_3$ and $\tau_4$ are assumed to take the same value $\tau$ for the sake of simplicity.  These functions imply that the interaction among the neurons in the first layer is excitatory in the short range but the effective interaction  becomes inhibitory in the long range via the inhibitory neurons in the second layer. The uniform inputs are assumed to be $I_{01}=15$ and $I_{02}=0$, that is, the neurons in the second layer do not oscillate without the interaction with the first layer. 
Figure 8(a) displays the average linear growth rate $\lambda_k$ as a function of $k$ for $\tau=0.05$, 1, 3 and 4.  The behavior of the growth rate is more complicated, compared to the previous one-layer models. The linear growth rates  are the largest at $\tau=1$, that is, the synchronized state is most unstable for the intermediate response time. The wave number for the most unstable mode decreases as $\tau$ is increased from 1 to 3, and becomes 0 for $\tau\sim 3.2$. 
Figure 8(b) displays the maximum value of $\lambda$ and $\lambda_{\pi}$ for the wavenumber $k=\pi$ as a function of $\tau$. The maximum growth rate increases with $\tau$ in the small range of $\tau$,  and then decreases for $\tau>1.2$, and becomes 0 for $\tau>3.2$.  The uniform state is stable for $\tau>3.2$. Inhibitory interaction is dominant in this coupling function, because $\int_{-\infty}^{\infty}[g_1(x)+g_2(x)]dx$ is negative; therefore the larger response time makes the synchronized motion more stable. 
The linear growth rate for $k=\pi$ is positive for $0.61<\tau<2.1$.
 However, the critical response time $\tau_c=3.2$ is rather smaller compared to the critical response time $\tau=6.4$ of the one-layer model as shown in Figs.~6 and 7. This means that the two-layer system is favorable for the synchronized motion, which is consistent with previous studies \cite{rf:8}.
Figure 8(c) displays a raster plot of firing neurons in the first layer at $\tau=0.05$.  Clustering domains appear, and the phase difference between neighboring clustering domains is nearly $\pi$. Most neighboring neurons belongs to the same domain and discontinuities appear only between the neighboring domains, in contrast to the purely inhibitory system shown in Fig.~5.  The sizes of the clustering domains are randomly distributed, and they depend on the initial conditions,  but they are roughly determined by the wavelength $l\sim 2\pi/0.06\sim 105$ of the most unstable Fourier mode in Fig.~8(a).  
Figure 8(d) displays a raster plot of firing neurons in the first layer at $\tau=2$. A traveling wave state with two pacemakers appears.   

We have investigated another two-layer model with $g_1(x)=1.35\exp(-0.03|x|), g_2(x)=-3.6\exp(-0.12|x|),g_3(x)=3.6\exp(-0.12|x|)$, and $g_4(x)=0$.
For this coupling, the interaction among the neurons in the first layer is excitatory in the long range, but the effective interaction is inhibitory in the short range via the inhibitory neurons in the second layer. The excitatory coupling is dominant in this model, because $\int_{-\infty}^{\infty}[g_1(x)+g_2(x)]dx$ is positive. 
Figure 9(a) displays the average linear growth rate $\lambda_k$ for $\tau=22,18$ and 16. At $\tau=16$, the synchronized state is stable. At $\tau=18$, the linear growth rate is weakly positive only for very small wave number. At $\tau=22$, 
the linear growth rate is positive for all wave numbers. The weak instability for very small wave number occurs at $\tau=16.9$ and the linear growth rates for all wave numbers become positive at $\tau=21.7$. In any case, the critical response time is rather larger than the case of the one-layer model shown in Fig.~1. This is another example where the interaction via inhibitory interneurons makes the synchronization more stable. Figure 9(b) displays a raster plot of the firing neurons in the first layer at $\tau=22$, where the short-wavelength instability occurs. The chimera state appears again also in this model as a result of the instability of the synchronized motion. 
\section{Summary}
We have studied the linear stability of the synchronized motion in the nonlocally coupled Hodgkin-Huxley equations.  We have found various  nonuniform oscillatory states  such as chimera states, wavy states with several pacemakers, antisynchronous clustering domains, and spatiotemporal chaos as a result of the instability of synchronized motion.  
We did not show numerical results, but qualitatively the same results were obtained for a coupling function of a sum of two Gaussian functions $g(x)=g_1\alpha_1/\sqrt{\pi}\exp[-(\alpha_1x^)2]+g_2\alpha_2/\sqrt{\pi}\exp[-(\alpha_2x)^2]$. The nonlocal coupling is essential for the appearance of discontinuities in the chimera states, clustering states, and spatiotemporal chaos with phase slips.
 
Some types of asynchronous states might be useful for information processing. It is important to make anomalous synchronization change into an asynchronous state in  the case of neurological diseases. We hope that our numerical results of various asynchronous states might be relevant to these applications.

\end{document}